\documentclass[12pt]{article}
\usepackage{setspace}
\usepackage{epsfig}
\usepackage{amssymb}
\def\be{\begin{equation}}
\def\ee{\end{equation}}
\def\ba{\begin{array}}
\def\ea{\end{array}}
\def\beqn{\begin{eqnarray}}
\def\eeqn{\end{eqnarray}}
\def\nonum{\nonumber}
\def\bt{\begin{tabular}}
\def\et{\end{tabular}}
\def\bc{\begin{center}}
\def\ec{\end{center}}
\begin{document}

\title{Implications of CP asymmetry parameter sin2$\beta$ on structural features of texture specific mass matrices}

\author{Rohit Verma$^1$, Gulsheen Ahuja$^2$, Manmohan Gupta$^2$\\
\\
 {$^1$ \it Rayat Institute of Engineering and Information Technology, Ropar, India.} \\
{$^2$ \it Department of Physics, Centre of Advanced Study, P.U.,
 Chandigarh, India.}\\
 {\it Email: mmgupta@pu.ac.in}}

 \maketitle
 \begin{abstract}
In the context of Fritzsch-like texture 4 zero Hermitian quark
mass matrices, we have attempted to find an `exact' formula for
sin$\,2\beta$ wherein the dependence of $\beta$ on the quark
masses and the elements of the quark mass matrices is visible in a
simple and clear manner. This has been achieved keeping in mind
the strong hierarchy of the quark masses and assuming the weak
hierarchy amongst the elements of the mass matrices. This `exact'
formula represents a vast improvement over the leading order
formula based on strong hierarchy of the elements of the mass
matrices. Apart from showing the compatibility of texture 4 zero
mass matrices with the present value of sin$\,2\beta$ and other
CKM parameters, we find interesting conclusions regarding the
structural features of the mass matrices.
 \end{abstract}
~~~~~~~~Keywords: Mass matrices, Quark mixing, CP asymmetry
parameter sin$\,2\beta$.

 ~~~PACS numbers: 12.15.Ff, 11.30.Er
\\
\section{Introduction}
The last few years have seen a precise measurement of
sin$\,2\beta$, characterizing CP asymmetry $a_{\psi K_s}$ in the
$B^o_d \rightarrow \psi K_s$ decay, as well as of other well
measured Cabibbo-Kobayashi-Maskawa (CKM) \cite{ckm} matrix
elements $V_{us}, V_{cb}, V_{ud}$. Based on these precise values,
several phenomenological analyses \cite{pdg08}-\cite{hfag} have
allowed us to conclude that the single CKM phase looks to be a
viable solution of CP violation not only in the case of K-decays
but also in the context of B-decays, at least to the leading
order. Interestingly, it has also been shown \cite{ouruni} that
the present value of sin$\,2\beta$ along with unitarity and other
well measured CKM parameters leads to an almost precise value of
$V_{ub}$ and CP violating phase $\delta$, two important parameters
yet to be measured precisely. Keeping in mind that the parameter
sin$\,2\beta$ also provides vital clues to the structural features
of texture specific mass matrices, comprising of hierarchy and
phases of the elements of the mass matrices, several authors
\cite{hallraisin}-\cite{bando} have explored its implications for
these. In particular, using assumption of `strong hierarchy' of
the elements of the mass matrix, having its motivation in the
hierarchy of the quark mixing angles, the following leading order
relationships between the various elements of the mixing matrix
and quark masses have been obtained in
\cite{hallraisin}-\cite{branco}, \beqn \left|
{\frac{V_{ub}}{V_{cb}}} \right| =
\sqrt{\frac{m_u}{m_c}},~~~~~~~~\left| {\frac{V_{td}}{V_{ts}}}
\right| = \sqrt{\frac{m_d}{m_s}},~~~~~~~\left| V_{us} \right| =
\left| {\sqrt{\frac{m_d}{m_s}}}e^{i \phi} -
\sqrt{{\frac{m_u}{m_c}}} \right|. \label{ratios} \eeqn  Following
Particle Data Group (PDG) \cite{pdg08} definition, these further
give the expression for $\beta$ in the `strong hierarchy' case,
e.g.,
\begin{eqnarray} \beta\equiv{\rm arg}\left[-\frac{V_{cd} V_{cb}^*}{V_{td}
 V_{tb}^*}\right]={\rm arg}\left[ 1- \sqrt{\frac{m_u m_s}{m_c m_d}} e^{-i \phi}
  \right].\label{betaothers} \end{eqnarray}
Unfortunately, the value of sin$\,2\beta$ predicted by the above
formula is in quite disagreement with its present precisely known
value. In particular, with the present values of input quark
masses and by giving full variation to phase $\phi$, the maximum
value of sin$\,2\beta$ comes out to be 0.5, which is in sharp
conflict with its present PDG 2008 \cite{pdg08} value $0.681 \pm
0.025$. Attempts have been made to resolve this conflict
\cite{frixingt4s2b}- \cite{branco}, however without getting into
detailed and comprehensive analysis of the issues involved in the
formulation of the above equations. Further, it may be added that
this issue escapes explicit attention of some of the recent
analyses \cite{matsuda, xingzhang}. Therefore, a closer look at
the whole issue is very much desirable.

It may be further noted that Fritzsch-like texture 4 zero mass
matrices are known to be compatible with specific models of GUTs
\cite{bando, 9912358, 0307359}, Abelian family symmetries
\cite{abelian}, as well as describe the neutrino oscillation data
quite well \cite{ournu}. It may also be kept in mind that the
mixing patterns of quarks and neutrinos are quite different, e.g.,
in the case of neutrinos, neither the mixing angles nor the
neutrino masses show any hierarchy, this being in sharp contrast
to the distinct hierarchy shown by quark masses and mixing angles.
These distinct features of quark and neutrino mixings may have a
constraining effect on mass matrices, particularly in case if
these have a common origin due to quark lepton unification
hypothesis \cite{qlepuni}. In this context, the question of
compatibility of texture 4 zero Fritzsch-like mass matrices with
precisely known sin$\,2\beta$ warrants a closer scrutiny as it may
provide vital clues regarding the structural features, including
the hierarchy and the phase structure of the elements of the mass
matrices.

Before reaching at any firm conclusion in this regard, we have to
address several issues. A precisely known sin$\,2\beta$ perhaps
requires a careful look at the relations arrived at in equations
(\ref{ratios}) and (\ref{betaothers}), which were derived in the
absence of precise knowledge regarding the parameter
sin$\,2\beta$. It seems that while arriving at the above relations
one of the key assumption used is that hierarchical mixing angles
are reproduced by `strongly hierarchical' mass matrices, having
bearing on the structural features of the mass matrices. This
assumption requires a careful and detailed scrutiny in the era of
precision measurements of CKM parameters.

In this context, one may note that the hierarchy followed by
mixing angles is $s_{13} < s_{23} < s_{12}$, whereas the quark
masses follow a somewhat stronger hierarchy, e.g., $m_u \ll m_c
\ll m_t$ and $m_d
< m_s \ll m_b$. The situation gets further complicated when one
notes that the mixing angles are not proportional to the quark
masses, in fact as can be seen from equation (\ref{ratios}), they
involve square roots of the ratios of the masses with certain
phase factors between the up and down quark sectors. This suggests
that in case we have to deal with a precisely known sin$\,2\beta$,
then one has to be careful in invoking a particular hierarchy
between the elements of the mass matrices. Further, in view of the
dependence of sin$\,2\beta$ on the ratios of the small quark
masses, it becomes essential to find an exact relationship
involving  small quark masses as well as the phases involved in
the mass matrices which have bearing on the hierarchy of the
elements of the mass matrices. Furthermore, one may wonder whether
only one phase is sufficient to support the data or one requires
two phases as are available in the case of texture 4 zero
Fritzsch-like Hermitian mass matrices.

From an analysis of texture 4 zero mass matrices it is very easy
to check that the issues raised above are not easy to answer,
e.g., in case we relax the `strong hierarchy' conditions on the
elements of the mass matrices then this immediately makes the task
of relating the CKM matrix elements to the ratio of masses, given
in equation (\ref{ratios}), a complicated affair. In principle, it
is an easy task to diagonalize exactly the texture 4 zero
Fritzsch-like Hermitian mass matrices, however in practice the
expressions of the CKM matrix elements so obtained are quite
lengthy, resulting in an expression for $\beta$ having complicated
dependence on quark masses, phases and free parameters of quark
mass matrices. Such an expression, although exact, does not allow
an insight into the role of phases, hierarchy of the elements of
the mass matrices or on the contributions of the non leading
terms. The first step in this direction is to develop a formula
for sin$\,2\beta$ which allows one to not only go beyond the
leading order but also to study the structural features, e.g., the
phases and the hierarchy of the elements of the mass matrices.

The purpose of the present paper on the one hand is to develop an
exact expression for sin$\,2\beta$ in terms of quark masses,
phases and free parameters of the texture specific mass matrices.
On the other hand, we would like to study, in detail, the
implications of such a formula on the compatibility of texture 4
zero mass matrices with sin$\,2\beta$. In particular, keeping in
mind the recent refinements of the  CKM matrix elements, we would
like to investigate in detail the implications of the exact
formula on the structural features of the mass matrices such as
the hierarchy of the elements of the mass matrices and their phase
structures. We would also like to examine the relationship between
the earlier formula of sin$\,2\beta$ and the present one derived
here.

The detailed plan of the paper is as follows. In Section
(\ref{form}), we detail the essentials of the formalism regarding
the texture specific mass matrices as well as the derivation of
the formula for sin$\,2\beta$. Inputs used in the present analysis
and the discussion of the calculations and results have been given
in Section (\ref{cal}). Finally, Section (\ref{summ}) summarizes
our conclusions.

\section{Texture specific mass matrices and the formula for sin$\,2\beta$\label{form}}
To fix the notations and conventions as well as to facilitate the
understanding of the relationship of the present work with the
earlier attempts, we detail some of the essentials of the
formalism. To begin with, we define the modified Fritzsch-like
matrices, e.g., \be M_i = \left( \ba {ccc} 0 & A_i & 0 \\ A_i^{*}
& D_i & B_i
\\
            0 & B_i^{*} & C_i \ea \right), \qquad i=U,D\,, \label{uniq} \ee
$M_U$ and $M_D$, respectively corresponding to the mass matrix in
the up sector and the down sector. It may be noted that each of
the above matrix is texture 2 zero type with $A_{i}
=|A_{i}|e^{i\alpha_{i}}$ and $B_{i} = |B_{i}|e^{i\beta_{i}}$. The
phases of the elements of the mass matrices $A_{i}$, $B_{i}$,
$C_{i}$, $D_{i}$ and their relative magnitudes characterize the
structural features of the mass matrices. A strongly hierarchical
mass matrix would imply $|A_{i}| \ll D_i \lesssim |B_{i}| < C_i$,
whereas a weaker hierarchy of the mass matrix implies $|A_{i}|
< D_{i} \lesssim |B_i| \lesssim C_i$. For the purpose of numerical work, one can
conveniently take the ratio $D_i/C_i \sim 0.01$ characterizing
strong hierarchy whereas $D_i/C_i \gtrsim 0.1$ implying weak
hierarchy. This can be understood by expressing these parameters
in terms of the quark masses, in particular $D_U/C_U \sim 0.01$
implies $C_U \sim m_t$ and $D_D/C_D \sim 0.01$ leads to $C_D \sim
m_b$.

The mass matrices $M_U$ and $M_D$ can be exactly diagonalized, for
details we refer the reader to \cite{groupmm} . To facilitate
diagonalization, the complex matrix $M_i$ can be expressed in
terms of the real matrix $M_i^r$ which can be diagonalized by the
orthogonal transformation, for example, \be M_i^{\rm diag} = O_i^T
M_i^{r} O_i
 \,,   \label{o1}\ee
where \be M_i^{\rm diag} = {\rm diag}(m_1,\,-m_2,\,m_3)\,, \ee the
subscripts 1, 2 and 3 referring respectively to $u,\, c$ and $t$
for the up sector and $d,\,s$ and $b$ for the down sector. The
negative sign before $m_2$ is only for the convenience of
calculations, without having physical significance.

The exact diagonalizing transformation $O_i$ is expressed as
 \be O_i
= \left( {\renewcommand{\arraystretch}{1.7}
 \ba{ccc}
  {\sqrt \frac{m_2 m_3 (C_i-m_1)}{(m_3-m_1)(m_2+m_1)C_i} } &
   {\sqrt \frac{m_1 m_3 (C_i+m_2)}{C_i (m_2+m_1) (m_3+m_2)}} &
  {\sqrt \frac{m_1 m_2 (m_3-C_i)}{C_i (m_3+m_2)(m_3-m_1)}}\\
 {\sqrt \frac{m_1 (C_i-m_1)}{(m_3-m_1)(m_2+m_1)} } &
 -{\sqrt \frac{m_2 (C_i+m_2)}{(m_3+m_2)(m_2+m_1)} }&
 {\sqrt \frac{m_3(m_3-C_i)}{(m_3+m_2)(m_3-m_1)} } \\
 -{\sqrt \frac{m_1 (m_3-C_i)(C_i+m_2)}{C_i(m_3-m_1)(m_2+m_1)} } &
 {\sqrt \frac{m_2 (C_i-m_1) (m_3-C_i)}{C_i
(m_3+m_2)(m_2+m_1)} } &
  {\sqrt \frac{m_3 (C_i-m_1)(C_i+m_2)}{C_i
(m_3+m_2)(m_3-m_1)}}  \ea} \right). \label{ou} \ee \vskip 0.5cm It
may be noted that while finding the above diagonalizing
transformation $O_i$ one has the freedom to choose several
equivalent possibilities of phases. Similarly, while normalizing
the diagonalized matrix to quark masses, one again has the freedom
to choose the phases for the quark masses. Out of these several
possibilities, we arrive at the above mentioned expression for
$O_i$ by considering the phase of $m_2$ to be negative,
facilitating the diagonalization process as well as the
construction of the CKM matrix.

The CKM mixing matrix $V_{\rm CKM}$ which measures the non-trivial
mismatch between diagonalizations of $M_U$ and $M_D$ can be
obtained using $O_{U(D)}$ through the relation
\be
V_{\rm CKM}= O_U^T (P_U P_D^{\dagger}) O_D. \ee Explicitly, the
elements of the CKM mixing matrix can be expressed as
\be
V_{l m} = O_{1 l}^U O_{1 m}^D e^{-i \phi_1} + O_{2 l}^U O_{2 m}^D
 + O_{3 l}^U O_{3 m}^D e^{i \phi_2},
\label{vckmelement} \ee
 where the subscripts $l$ and $m$ run respectively over $u,\, c$, $t$  and $d,\,s$, $b$
and $\phi_1=  \alpha_U- \alpha_D$, $\phi_2= \beta_U- \beta_D$.

Using the above equation, the elements of the CKM mixing matrix
can be easily found, e.g.,
 \beqn
  V_{cd}=\sqrt{\frac{m_u m_t (-D_u + m_u +
m_t)}{C_u(m_u+m_c)(m_c+m_t)}} \sqrt{\frac{m_s m_b(-D_d + m_b -
m_s)}{C_d(m_b-m_d)(m_s+m_d)}}~e^{-i \phi_1}  \nonum
\\- \sqrt{\frac{m_c (-D_{u} + m_u +
m_t)}{(m_u+m_c)(m_c+m_t)}} \sqrt{\frac{m_d (-D_d + m_b -
m_s)}{(m_b-m_d)(m_s+m_d)}}~~~~~~~~~~~~~~~~~~~ \nonum
\\ - \sqrt{\frac{m_c (-D_u + m_t - m_c)(D_u - m_u +
m_c)}{C_u(m_u+m_c)(m_t+m_c)}} \times
~~~~~~~~~~~~~~~~~~~~~~~~~~\nonum
\\ \sqrt{\frac{m_d (D_d - m_d + m_s)(-D_d + m_d +
m_b)}{C_d(m_b-m_d)(m_s+m_d)}}~e^{i
\phi_2}.~~~~~~~~~~~~~~~~~~~~~~~~~\label{fvcd}
  \eeqn
 The other elements $ V_{cb}$, $V_{td}$ and $V_{tb}$, also required to be known
to evaluate $\beta$, can also be obtained similarly. In case we
use the above complicated expression for $V_{cd}$ as well as
similar expressions of the other elements to evaluate
sin$\,2\beta$, we find that these would yield a long and
complicated formula from which it would be difficult to understand
the implications on the phases and other parameters of the mass
matrices. To derive a simple and informative formula, we first
rewrite the diagonalizing transformation $O_i$ keeping in mind
$m_3 \gg m_2 \gg m_1$ and the element of the mass matrix $C_i \gg
m_1$, which is always valid without any dependence on the
hierarchy of the elements of the mass matrices. It may be
mentioned that this approximation induces less than a fraction of
a percentage error in the numerical results. The structure of
$O_i$ can be simplified and expressed as
 \be O_i
= \left( {\renewcommand{\arraystretch}{1.7}
 \ba{ccc}
  1&
   \zeta_{1 i} {\sqrt \frac{m_1}{m_2}} &
 {\frac{\zeta_{2 i}}{\zeta_{3 i}}} {\sqrt \frac{m_1 m_2}{m_3^2}} \\
  \zeta_{3 i} {\sqrt \frac{m_1}{m_2}} &
  - \zeta_{1 i} \zeta_{3 i}&
 \zeta_{2 i} \\
  - \zeta_{1 i} \zeta_{2 i} {\sqrt \frac{m_1}{m_2}} &
  \zeta_{2 i} &
  \zeta_{1 i} \zeta_{3 i} \ea} \right), \label{oisimple2} \ee
  where the three parameters $\zeta_{1 i}$, $\zeta_{2 i}$, $\zeta_{3
i}$, with $i$ denoting $U$ and $D$ are given by \be \zeta_{1 i}=
\sqrt{1 + \frac{m_2}{C_i}},~~~~\zeta_{2 i}= \sqrt{1 -
\frac{C_i}{m_3}},~~~~\zeta_{3 i}= \sqrt{\frac{C_i}{m_3}}. \ee
Making use of this equation, along with relation
(\ref{vckmelement}), we obtain the following elements needed to
evaluate $\beta$
 \be V_{cd} = \zeta_{1 U} \sqrt{\frac{m_u}{m_c}}
e^{-i \phi_1} - \sqrt{\frac{m_d}{m_s}}~ \left[ \zeta_{1 U}\,
\zeta_{3 U}\, \zeta_{3 D} + \zeta_{2 U}\, \zeta_{1 D}\, \zeta_{2
D}\, e^{i \phi_2} \right], \label{cvcd}\ee

 \be V_{cb} = \frac{\zeta_{1 U} \zeta_{2 D}}{\zeta_{3 D}} \sqrt{\frac{m_u m_d m_s}{m_c m_b^2}}
e^{-i \phi_1} - \left[ \zeta_{1 U}\, \zeta_{3 U}\, \zeta_{2 D} -
\zeta_{2 U}\, \zeta_{1 D}\, \zeta_{3 D}\, e^{i \phi_2} \right],
\label{cvcb}\ee

 \be V_{td} = \frac{\zeta_{2 U}}{\zeta_{3 U}} \sqrt{\frac{m_u m_c}{m_t^2}}
e^{-i \phi_1} + \sqrt{\frac{m_d}{m_s}}~\left[ \zeta_{2 U}\,
\zeta_{3 D} - \zeta_{1 U}\, \zeta_{3 U}\, \zeta_{1 D}\, \zeta_{2
D}\, e^{i \phi_2} \right], \label{cvtd}\ee

\be V_{tb} = \frac{\zeta_{2 U} \zeta_{2 D}}{\zeta_{3 U} \zeta_{3
D}} \sqrt{\frac{m_u m_c m_d m_s}{m_t^2 m_b^2}} e^{-i \phi_1} +
\left[ \zeta_{2 U}\, \zeta_{2 D} + \zeta_{1 U}\, \zeta_{3 U}\,
\zeta_{1 D}\, \zeta_{3 D}\, e^{i \phi_2} \right]. \label{cvtb}\ee
A general look on the above elements clearly shows that the above
relations are not only more compact but also more useful to view
the dependence of these CKM matrix elements on the quark masses
and phases. Using these elements, after some non trivial algebra,
one arrives at the following expression of $\beta$, wherein its
dependence on the quark masses and the elements of the quark mass
matrices is visible in a simple and clear manner, e.g.,
\begin{eqnarray} \beta\equiv{\rm arg}\left[-\frac{V_{cd}
V_{cb}^*}{V_{td}
 V_{tb}^*}\right]={\rm arg}\left[ \left( 1- \sqrt{\frac{m_u m_s}{m_c m_d}} e^{-i (\phi_1 + \phi_2)} \right)
 \left( \frac{1-r_2 e^{i \phi_2}}{1-r_1 e^{i \phi_2}} \right)  \right]
,\label{beta} \end{eqnarray} where the parameters $r_1$ and $r_2$
can be expressed in terms of the quark masses and the elements of
the quark mass matrices via the relations,
 \be r_1 = \frac{\zeta_{1 U} \,
\zeta_{3 U}\, \zeta_{1 D}\, \zeta_{2 D}}{\zeta_{2 U}\, \zeta_{3
D}}~~~~~~~~~{\rm and}~~~~~~~~~ r_2 = \frac{\zeta_{1 U} \, \zeta_{3
U}\, \zeta_{2 D}}{\zeta_{2 U}\, \zeta_{1 D}\, \zeta_{3 D}}. \ee

The relationship derived by us, given in equation (\ref{beta}), is
an `exact' formula emanating from texture 4 zero mass matrices,
incorporating both the phases. This formula has several
interesting aspects. Apart from clearly underlying the dependence
of small quark masses and the phases $\phi_1$ and $\phi_2$, it
also clearly establishes the modification of the earlier formula.
It is very easy to check that the earlier formula can be easily
deduced from the present one by using the strong hierarchy
assumption which essentially translates to $\zeta_{1 D} \simeq
\zeta_{1 U} \simeq 1$, further implying $r_{1}$=$r_{2}$, leading
to the term $\left( \frac{1-r_2 e^{i \phi_2}}{1-r_1 e^{i \phi_2}}
\right)$ becoming 1. The phase of the earlier formula can be
obtained by identifying $\phi_1 + \phi_2$ as $\phi$, taking values
from 0 to 2$\pi$.

It also needs to be re-emphasized that while arriving at the above
`exact' relationship, we have considered the hierarchy of the
quark masses, e.g., $m_t \gg m_u$ and $m_b \gg m_d$ as well as
have used $m_3 \gg m_2 \gg m_1$ and the element of the mass matrix
$C_i \gg m_1$, these being valid in both weak and strong hierarchy
cases. It may also be added that the formula remains valid for
both the weak hierarchy of the elements of the mass matrices given
by $|A_{i}|< D_{i} \lesssim |B_i| \lesssim C_i$ as well as for the
strong hierarchy assumption $|A_{i}| \ll D_i \lesssim |B_{i}| <
C_i$. Interestingly, the modification to the earlier formula
contributes only when $\phi_2 \neq 0$, implying thereby that both
the phases of the mass matrices might play an important role in
achieving the agreement with data.

\section{Calculations and results\label{cal}}
In order to investigate the implications of the formula given in
equation (\ref{beta}) on the structural features of the mass
matrices and the CKM parameters, as a first step we find the range
of sin$\,2\beta$ predicted by the above formula by using the
latest inputs. In this regard, we have adopted the following
ranges of quark masses \cite{xinginput} at the energy scale of
$M_z$, e.g., \beqn m_u=0.8 -1.8\, {\rm MeV},~~ m_d=1.7 -4.2\, {\rm
MeV},~~ m_s=40.0 -71.0\, {\rm MeV},~~~~~~~~\nonumber\\
~~~~~~~m_c=0.6- 0.7\, {\rm GeV},~~ m_b=2.8- 3.0\, {\rm GeV},~~
m_t=169.5- 175.5\, {\rm GeV}. ~~~~~~~~\label{qmasses} \eeqn With
these inputs and the `exact' formula given in equation
(\ref{beta}), we have evaluated sin$\,2\beta$ by giving full
variation to the phases $\phi_1$ and $\phi_2$, the parameters
$D_U$ and $D_D$ have been given wide variation in conformity with
the natural hierarchy of the elements of the mass matrices e.g.,
$D_i < C_i$ for $i=U, D$. The sin$\,2\beta$ so evaluated comes out
to be
\be
\\{\rm sin}\,2\beta = 0.4105 - 0.7331.\ee Interestingly, we find that
the above value is inclusive of its experimental range $0.681\pm
0.025$. This clearly indicates that the exact formula which
includes weak hierarchy as well as additional phase factors plays
a crucial role in bringing out reconciliation between texture 4
zero mass matrices and the present precise value of sin$\,2\beta$.

Before we get into examining the detailed implications of
sin$\,2\beta$ on structural features of the mass matrices, it is
perhaps desirable to check the compatibility of texture 4 zero
mass matrices with other precisely known parameters of the CKM
phenomenology. To this end, along with the latest experimental
value of sin$\,2\beta$, we have imposed the following PDG 2008
\cite{pdg08} constraints given by \be |V_{us}|=0.2255\pm
0.0019,~~~~|V_{cb}|=(41.2\pm 1.1) 10^{-3},~~~~ |V_{ub}|= 0.0035\pm
0.0002.\ee As mentioned earlier, we have considered the value of
$|V_{ub}|$ obtained recently \cite{ouruni} using only the
unitarity of the elements of the CKM matrix and current
sin$\,2\beta$ value. Keeping in mind the above mentioned
constraints and using the current values of quark masses given in
equation (\ref{qmasses}), we have evaluated the entire CKM matrix
at 1$\sigma$ C.L. by using equations (\ref{cvcd})-(\ref{cvtb}) and
the other corresponding expressions for the remaining CKM matrix
elements, e.g.,
 \be V_{{\rm CKM}} = \left( \ba{ccc}
   0.9738- 0.9747 &   0.2236 - 0.2274 &  0.0033 - 0.0037 \\
 0.2234 - 0.2273  &   0.9729 - 0.9739  &  0.0401 - 0.0423\\
0.0068 - 0.0103  &  0.0390 - 0.0417 &  0.9991 - 0.9992 \ea
\right). \label{1sm} \ee A general look at the matrix reveals that
the ranges of CKM elements obtained here are quite compatible with
those obtained by recent global analyses. In particular, the
ranges found here are in good agreement with those emerging from
global fits by PDG 2008 \cite{pdg08}, UTfit \cite{utfit},
CKMfitter \cite{ckmfitter} and HFAG \cite{hfag}.

\begin{figure}
\psfig{file=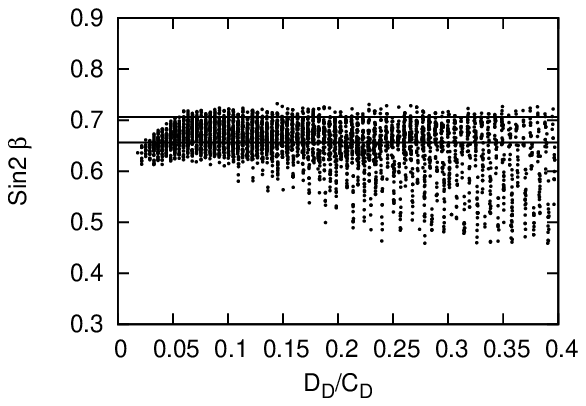, width=4.5in} \caption{Plot showing
variation of CP violating parameter sin$\,2\beta$ versus hierarchy
characterizing ratio $D_D/C_D$}
  \label{s2bddcd}
  \end{figure}

After having shown the compatibility of texture 4 zero mass
matrices with sin$\,2\beta$ as well as the recent ranges of the
elements of the CKM matrix, we investigate the implications of the
formula derived in equation (\ref{beta}) on the structural
features of mass matrices. In particular, we examine the
constraints imposed on the ratio $D_i/C_i$ for $i=U, D$,
characterizing hierarchy, as well as on the phases $\phi_1$ and
$\phi_2$ of the mass matrices. As a first step, using exact
relation obtained earlier, we investigate the role of hierarchy by
plotting sin$\,2\beta$ against the ratio $D_D/C_D$ in figure
\ref{s2bddcd}. Several interesting conclusions follow from the
graph. It can be easily noted that when $D_D/C_D$ $<$ 0.02, we are
not able to reproduce any point within the 1$\sigma$ range of
sin$\,2\beta$, even after giving full variation to all the other
parameters. It may be of interest to mention that the earlier
attempts \cite{hallraisin}-\cite{branco} had considered a value of
$D_D/C_D$ $\lesssim$ 0.02, thereby resulting in the
incompatibility of texture 4 zero mass matrices with
sin$\,2\beta$. From the figure it can be easily checked that only
for $D_D/C_D$ $>$ 0.05, full range of sin$\,2\beta$ is reproduced.
This clearly shows that as we deviate from strong hierarchy
characterized by the ratio $D_D/C_D \sim 0.01$ towards weak
hierarchy given by $D_D/C_D \gtrsim 0.1$, we are able to reproduce
the results. It may be mentioned that although the graph has been
plotted for $D_D/C_D$ up to 0.4, however the same pattern is
followed up to $D_D/C_D \sim 0.6$, beyond which the basic
structure of the mass matrix is changed. It may be added that the
corresponding graph of $D_U/C_U$ is also very much similar. One
would also like to emphasize that the agreement between
sin$\,2\beta$ and higher values of $D_i/C_i$ does not spoil the
overall agreement of texture 4 zero mass matrices with the CKM
matrix derived earlier. This brings out an extremely important
point as the conventional belief was that the hierarchical quark
mixing angles can be reproduced only by strong hierarchy mass
matrices. Further, we believe that this point would provide strong
impetus for quark-lepton unification at the GUTs scale.

\begin{figure}
\psfig{file=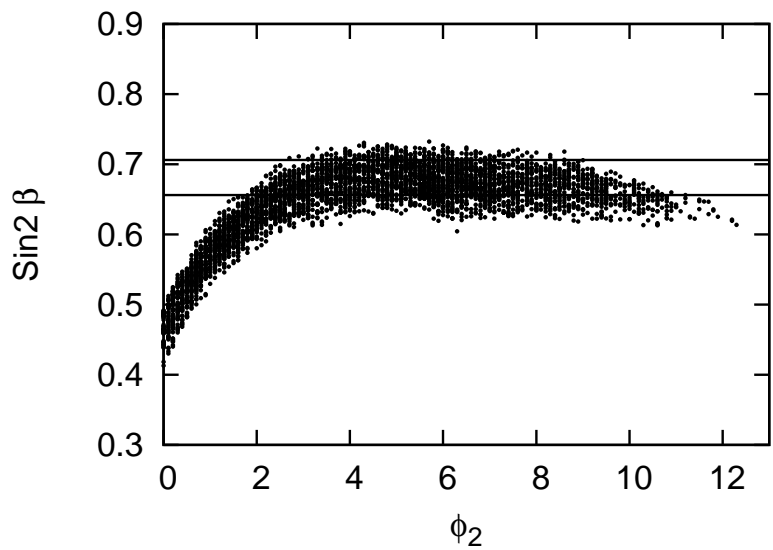, width=4.5in} \caption{Plot showing
variation of CP violating parameter sin$\,2\beta$ versus the phase
$\phi_2$}
  \label{s2bph2}
  \end{figure}

Coming to the issue of phases $\phi_1$ and $\phi_2$ of the mass
matrices, it may be pointed out that our analysis yields $\phi_1$
from $70^{\rm o}$ - $90^{\rm o}$ and $\phi_2$ takes values from
$3^{\rm o}$ - $9^{\rm o}$ in order to achieve compatibility with
the CKM matrix derived earlier. Most of the earlier analyses
considered only one phase $\phi_1$ whereas the phase $\phi_2$ was
assumed to be zero. However, as already mentioned, considering
phase $\phi_2$ to be zero immediately leads to incompatibility of
texture 4 zero mass matrices with sin$\,2\beta$, therefore a
detailed analysis pertaining to the role of phase $\phi_2$ is in
order. To this end, in figure \ref{s2bph2} we have plotted a graph
of sin$\,2\beta$ versus angle $\phi_2$. A close look at the graph
reveals several interesting points. In particular, the graph
clearly illustrates the crucial role played by the phase $\phi_2$
in bringing out agreement of texture 4 zero mass matrices and the
precisely known sin$\,2\beta$. It is interesting to emphasize that
despite giving full variation to other parameters, for
$\phi_2=0^{\rm o}$ we are not able to reproduce sin$\,2\beta$
within the experimental range.

After having realized the role of weak hierarchy, characterized by
the ratio $D_i/C_i$, as well as the phase $\phi_2$ being non zero,
for describing the present value of sin$\,2\beta$, we attempt to
assess the quantitative role of these parameters in improving its
value. To this end, we re-express $\beta$ as
\be
\beta = \beta_1+ \beta_2, \ee with
\be
\beta_1 = {\rm tan^{-1}}\left( 1- \sqrt{\frac{m_u m_s}{m_c m_d}}
e^{-i (\phi_1 + \phi_2)} \right),\beta_2 = {\rm tan^{-1}} \left(
\frac{(\zeta_{1 D}^2 -1)r_2 \,{\rm sin}\, \phi_2}{1+\zeta_{1 D}^2
r_2^2-(\zeta_{1 D}^2 +1)r_2 \,{\rm cos}\, \phi_2} \right).
\label{beta2} \ee It may be noted that $\beta_1$ corresponds to
the contribution given by the expression (\ref{betaothers}), with
$\phi_1 + \phi_2$ being identified as $\phi$, whereas $\beta_2$
represents additional contribution coming from retention of non
leading terms as well as due to weak hierarchy and additional
phase factors.

In Table \ref{tabs2bvdd/cd}, corresponding to the phase $\phi_2$
values from $0^{\rm o}- 12^{\rm o}$, we have presented values of
sin$\,2\beta$, $\beta_1$ and $\beta_2$ for some typical values of
$D_D/C_D$ to illustrate the role of these in achieving a
quantitative fit. It may be mentioned that the purpose here is not
to give a systematic interdependence of various parameters, rather
to give an idea about the amount of contribution of phase $\phi_2$
and weak hierarchy towards sin$\,2\beta$. From the table, one
finds that for $\phi_2=0^{\rm o}$ the corresponding value of
$\beta_2$ is zero, leading to a small value of sin$\,2\beta$.
However, as $\phi_2$ increases up to $\sim 3^{\rm o}$, the
corresponding values of sin$\,2\beta$ are within experimental
limits. As $\phi_2$ increases further up to $\sim 10^{\rm o}$, one
finds that sin$\,2\beta$ values still remain within the
experimental range, this being in agreement with the observations
of figure \ref{s2bph2}. One may wonder why a small change in phase
$\phi_2$ leads to a relatively large contribution to
sin$\,2\beta$. This can be understood from the exact relationship
between the parameter $\beta_2$ and the phase $\phi_2$ given in
equation (\ref{beta2}), showing that $\beta_2$ is represented by
ratio of two very small numbers, signifying that even a small
change in the value of $\phi_2$ can produce reasonable
contribution to $\beta_2$. Further, from the table some light is
also shed on the relative importance of hierarchy and the phase
$\phi_2$. In particular, one finds that corresponding to lower
values of the phase $\phi_2$, the ratio $D_D/C_D$ acquires
somewhat higher values as compared to the ones obtained by
increasing $\phi_2$.

\section{Summary and conclusions\label{summ}}
To summarize, in the context of Fritzsch-like texture 4 zero
Hermitian quark mass matrices, we have found an `exact' formula
for sin$\,2\beta$ wherein the dependence of $\beta$ on the quark
masses and the elements of the quark mass matrices is visible in a
simple and clear manner. This has been done keeping in mind $m_3
\gg m_2 \gg m_1$ and $C_i \gg m_1$ as well as the weak hierarchy
of the elements of the mass matrices given by $|A_{i}|
< D_{i} \lesssim |B_i| \lesssim C_i$. The `exact' formula found here represents a vast
improvement over the usual formula based on strong hierarchy of
the elements of the mass matrices. Besides clearly underlying the
compatibility of texture 4 zero mass matrices in the case of weak
hierarchy, the formula also explains why in the strong hierarchy
case we are unable to obtain the value of sin$\,2\beta$.

A detailed analysis based on the present formula as well as by
using other well measured CKM matrix elements shows that the
texture 4 zero Hermitian  mass matrices are compatible with recent
results emerging from global fits by PDG 2008 \cite{pdg08}, UTfit
\cite{utfit}, CKMfitter \cite{ckmfitter} and HFAG \cite{hfag}.
Further, the formula clearly provides a detailed insight into the
phase structure and the hierarchy of the elements of the mass
matrices. In fact, we find that both the phases $\phi_1$ and
$\phi_2$ are required to fit the data, with $\phi_1$ ranging from
$70^{\rm o}$ - $90^{\rm o}$, whereas $\phi_2$ taking values from
$3^{\rm o}$ - $9^{\rm o}$. In conclusion, we would like to state
that we can reproduce hierarchical mixing angles even with weakly
hierarchical mass matrices which may have vital implications for
quark-lepton unification hypothesis.

\vskip 0.5cm {\bf Acknowledgements} \\ RV would like to thank the
Director, RIEIT for providing facilities to work. GA would like to
thank DST, Government of India for financial support and the
Chairman, Department of Physics for providing facilities to work
in the department.

\newpage

  \begin{table}
\bc {\renewcommand{\arraystretch}{1.4}
\begin{tabular}{|c|c|c|c|c|} \hline
$~~\phi_2~~$ & $D_D/C_D$ & $\beta_1$ & $\beta_2$ & sin$\,2(\beta_1
+ \beta_2)$
\\ \hline
 0 & 0.6649 & 12.12 & 0.00 & 0.4106 \\
 1 & 0.3506 & 12.78 & 3.90 & 0.5499 \\
 2 & 0.2666 & 12.73 & 6.40 & 0.6192 \\
 3 & 0.2179 & 12.68 & 7.92 & 0.6587 \\
 4 & 0.1728 & 12.62 & 8.30 & 0.6671 \\
 5 & 0.1538 & 12.56 & 8.74 & 0.6769 \\
 6 & 0.1309 & 12.49 & 9.17 & 0.6861 \\
 7 & 0.1134 & 12.44 & 9.52 & 0.6937 \\
 8 & 0.0717 & 13.01 & 9.31 & 0.7026 \\
 9 & 0.0559 & 13.03 & 9.06 & 0.6969 \\
 10 & 0.0404 & 12.95 & 8.03 & 0.6686 \\
 11 & 0.0327 & 12.85 & 6.75 & 0.6320 \\
 12 & 0.0216 & 12.77 & 6.10 & 0.6121
\\ \hline
\end{tabular}} \ec
\caption{Some of the values of sin$\,2\beta$, $\beta_1$ and
$\beta_2$ obtained by varying $\phi_2$ from $0^{\rm o}- 12^{\rm
o}$. The angles $\beta_1$ and $\beta_2$ are in degrees.}
\label{tabs2bvdd/cd}
\end{table}


\begin{thebibliography}{99}
\bibitem{ckm}N. Cabibbo, Phys. Rev. Lett. 10, (1963) 531; M.
Kobayashi, T. Maskawa, Prog. Theor. Phys. 49, (1973) 652.

\bibitem{pdg08}C. Amsler {\it et al.},  Phys. Lett. B667, (2008) 1,
updated results available at \\ http://pdg.lbl.gov/.

\bibitem{utfit}M. Bona {\it et al.}, UTfit Collaboration, arXiv: 0803.0659,
arXiv: 0905.3747, updated results available at
http://www.utfit.org/.

\bibitem{ckmfitter}J. Charles {\it et al.}, CKMfitter Group, arXiv: 0905.1572,
updated results available at http://ckmfitter.in2p3.fr/.

\bibitem{hfag}E. Barberio {\it et al.}, Heavy Flavor Averaging Group (HFAG),
arXiv: 0808.1297, updated results available at
http://www.slac.stanford.edu/xorg/hfag/.

\bibitem{ouruni}G. Ahuja, M. Gupta, S. Kumar, M. Randhawa,
Phys. Lett. B647, (2007) 394.

\bibitem{hallraisin}L. J. Hall, A. Rasin, Phys. Lett. B315, (1993) 164.

\bibitem{rrr}P. Ramond, R. G. Roberts, G. G. Ross, Nucl. Phys. B406, (1993) 19.

\bibitem{barbieri}R. Barbieri, L. J. Hall, A. Romanino, Phys. Lett. B401, (1997)
47.

\bibitem{roberts}R. G. Roberts, A. Romanino, G. G. Ross,
L. Velasco-Sevilla, Nucl. Phys. B615, (2001) 358.

\bibitem{frixingt4s2b}H. Fritzsch, Z. Z. Xing, Phys. Lett. B555, (2003)
63.

\bibitem{kimraby}H. D. Kim, S. Raby, L. Schradin, Phys. Rev. D69, (2004)
092002.

\bibitem{branco}G. C. Branco, M. N. Rebelo, J. I. Silva-Marcos, Phys. Rev.
D76, (2007) 033008.

\bibitem{bando}M. Bando, S. Kaneko, M. Obara, M. Tanimoto, Prog. Theor. Phys.
116, (2007) 1105.

\bibitem{matsuda}K. Matsuda, H. Nishiura, Phys. Rev. D74, (2006) 033014.

\bibitem{xingzhang}Z. Z. Xing, H. Zhang, J. Phys. G30, (2004) 129.

\bibitem{9912358}H. Fritzsch, Z. Z. Xing,
Prog. Part. Nucl. Phys. 45, (2000) 1, and references therein.

\bibitem{0307359}Z. Z. Xing, Int. J. Mod. Phy. A19, (2004) 1, and
references therein.

\bibitem{abelian}W. Grimus, hep-ph/0511078.

\bibitem{ournu}M. Randhawa, G. Ahuja, M. Gupta, Phys. Rev. D65, (2002) 093016;
{\it ibid.} Phys. Lett. B643, (2006) 175; G. Ahuja, S. Kumar, M.
Randhawa, M. Gupta, S. Dev, Phys. Rev. D76, (2007) 013006.

\bibitem{qlepuni}A. Yu. Smirnov, hep-ph/0604213.

\bibitem{groupmm}P. S. Gill, M. Gupta, J. Phys. G: Nuc. Part.
Phys. 21, (1995) 1; M. Randhawa, V. Bhatnagar, P. S. Gill, M.
Gupta, Phys. Rev. D60, (1999) 051301.

\bibitem{xinginput}Z. Z. Xing, H. Zhang, S. Zhou, Phys. Rev. D77, (2008)
113016; H. Leutwyler, Phys. Lett. B378, (1996) 313.

\end{thebibliography}
\end{document}